\documentclass[12pt]{article}
\usepackage{amsmath,amssymb}
\usepackage{amscd}
\usepackage{epsfig}
\allowdisplaybreaks
\makeindex
\begin{document}
\title{Physics with exotic probability theory}
\author{%
  Saul Youssef%
  \hfil \\
  Department of Physics and \\
  Center for Computational Science \\
  Boston University \\
  youssef@bu.edu\\
}
\maketitle
\begin{abstract}
Probability theory can be modified in essentially one way while 
maintaining consistency with the basic Bayesian framework.  This
modification results in copies of standard probability theory for
real, complex or quaternion probabilities.  These copies, in turn, 
allow one to derive quantum theory while restoring standard probability 
theory in the classical limit.  The argument leading to these three
copies constrain physical theories in the same sense that Cox's original
arguments constrain alternatives to standard probability
theory.  This sequence is presented in some detail with emphasis 
on questions beyond basic quantum theory where new insights are needed.

\end{abstract}

\section{Introduction}

    If it weren't for the weight of history, it would seem 
natural to take quantum mechanical phenomena as an indication that
something has gone wrong with probability theory and to attempt to 
explain such phenomena by modifying probability theory itself, rather 
than by invoking quantum mechanics.  
It is actually easy to 
take this point of view because probability theory is so tightly 
constrained by Cox's Bayesian arguments\cite{cox} that there is only one 
plausible try.  Trying this anyway\cite{mpl2,pl,mpl1,santafe}, one finds that
Cox's arguments work even without
the assumption that probabilities are real and non--negative and
one obtains ``exotic'' copies of standard probability theory where the probabilities
may belong to any real associative algebra with unit.  With probability
theory modified, there is no need for the usual ``wave-particle duality''
and one is free to assume, for example, that a particle in ${\bf R}^3$ 
is somewhere in ${\bf R}^3$ at each time.  Introducing such ``state spaces''
and assuming that probabilities have a square norm, exotic probabilities
acquire the power to predict real non-negative frequencies and are 
limited to three algebras: reals, complex numbers and quaternions.
Given this framework, 
complex probabilities with state spaces ${\bf R}^3$ or ${\bf R^4}$ lead to 
the standard quantum theory in complete detail including the Schrodinger
equation and ``mixed states.'' Quaternionic probabilities lead,
on the other hand, to the Dirac 
theory\cite{srinivasan1,srinivasan2}.  
Although one might
expect such theories to be ruled out by Bell's arguments, modifying 
probability theory turns out to evade this and similar restrictions\cite{pl}.
Because of the simple
nature of the state space axioms and the Bayesian nature of the exotic 
probabilities, the familiar semi--paradoxical measurement and observer
questions from quantum theory do not arise\cite{santafe}.  One has a 
theory which is quite substantially simpler than quantum mechanics
both conceptually and mathematically.

    Although predictions within state spaces like ${\bf R}^3$ and ${\bf R}^4$ 
agree with standard quantum mechanics, 
Srinivasan has realized that one should expect even more interesting results in 
field theory because exotic probability theory cannot 
produce the apparent divergences which are so common in quantum field theory.  
Indeed, he has shown that with his quaternionic probability version of 
canonical quantization, he gets the correct result for the Lamb shift 
without any renormalization procedure\cite{srinivasan3}.

    This paper is intended as a review of the basic results from 
references 2-5 with more detail than is practical in 
letter sized papers, as a starting point for someone interested in this
general subject and as an exposition of unanswered questions
where further research is needed.  
The idea that probability theory might be altered in some way
goes back at least to Dirac\cite{dirac}.  For a history of 
this idea, the 
review by Muckenheim et al.\cite{muckenheimetal} is a good starting point.  
Related ideas can be found papers by Srinivasan and
Sudarshan\cite{srinivasan1,srinivasan2,srinivasan3},
Gudder\cite{gudder}, Feynman\cite{feynman},  Tikochinsky\cite{tikochinsky}, Frohner\cite{frohner},
Caticha\cite{caticha}, Steinberg\cite{steinberg}, Belinskii\cite{belinskii}, 
Miller\cite{miller}, Muckenheim\cite{muckenheim}, Khrennikov\cite{khrennikov} and Pitowsky\cite{pitowsky}.  This work is very  
influenced by the Bayesian view of probability theory due to 
Ed Jaynes\cite{jaynes1,jaynes2,jaynes3,ejaynes}.

\section{Cox arguments}

     In the Bayesian view of probability theory, probabilities begin as 
real non-negative numbers assigned to pairs $(a,b)$ of arbitrary propositions.
These numbers are meant to indicate, in some sense to be defined, how likely 
it is that proposition $b$ is true given that proposition $a$ is known. Given 
this setup,  Cox argued\cite{cox} that if such an assignment of numbers is to be useful as
 a likelihood, it should satisfy a few plausible conditions.  He then demonstrates 
(it is not a proof for reasons which will be clear below) that these conditions lead
unambiguously to the standard Bayesian presentation of probability theory.
The basic plan is to simply follow Cox's work while dropping the assumption
that probabilities are real and non-negative.  

    Before beginning, there are a couple of technical points which might 
cause confusion.  Cox\cite{cox} and Jaynes\cite{ejaynes} discuss probability theory without any
restriction on propositions. 
The idea is that probability theory is
meant to be ``the logic of science'' and is meant to be treated slightly 
informally in the same sense that ordinary logic is treated slightly 
informally in mathematics.  However, for definiteness, and since we will
introduce several copies of probability theory, we work in a distributive 
lattice.  The other technical point is that Cox, Jaynes and my previous 
papers work in a Boolean lattice as opposed to a distributive lattice.
It is easier to deal with a plain distributive lattice and this makes no 
difference for the results in references 2-5.

    Consider a set $P$ and a distributive lattice $L$ with ``propositions''
$a,b,c\in L$ with minimum element $0\in L$ and maximal element $1\in L$.
For a function $\rightarrow:L\times L\rightarrow P$ to be a useful 
measure of ``likelihood,'' 
we expect, following Cox\cite{cox}, that
$(a\rightarrow b)$ and $(a\wedge b\rightarrow c)$ should determine
$(a\rightarrow b\wedge c)$ and denote the implied function by
$*:P\times P\rightarrow P$.  Similarly, if $b\wedge c=0$, we also 
expect that $(a\rightarrow b)$ and $(a\rightarrow c)$ should determine
$(a\rightarrow b\wedge c)$ and denote this function by 
$+:P\times P\rightarrow P$. 
Mathematically speaking, Cox's point is that the structure of $L$
has implications for $*$ and $+$.  For example, for any $a,b,c,d\in L$,
we have 
\begin{equation}
(a\rightarrow b\wedge c\wedge d) = (a\rightarrow b)*(a\wedge b\rightarrow c\wedge d)
= (a\rightarrow b)*[(a\wedge b\rightarrow c)*(a\wedge b\wedge c\rightarrow d)]
\end{equation}
and using the associativity of $\wedge$,
\begin{equation}
(a\rightarrow b\wedge c\wedge d) = [(a\rightarrow b)*(a\wedge b\rightarrow c)]*(a\wedge b\wedge
c\rightarrow d).
\end{equation}
Letting $x=(a\rightarrow b)$, $y=(a\wedge b\rightarrow c)$ and $z=(a\wedge b\wedge c\rightarrow
d)$, we
have
\begin{equation}
x*(y*z) = (x*y)*z
\end{equation}
for all such triples $(x,y,z)$.  Following Cox, we further assume that $*$ is
associative in general.

   Similarly, suppose that we have $a,b,c\in L$ with $b\wedge c=0$.  Then
$(a\rightarrow b\vee c)=(a\rightarrow b)+(a\rightarrow c)=(a\rightarrow c)+(a\rightarrow b)$.
We then plausibly assume that $+$ is commutative in general.

    One can easily complete this picture checking properties of $L$ to see what
is correspondingly expected in $P$.

\begin{center}
    \begin{tabular}{|l|l|}
         Property of L & Expected property of P \\ \hline
         $\wedge$ is associative    & $*$ is associative \\ 
         $\vee$ is associative  & $+$ is associative \\
		 $\wedge$ is commutative & ------ \\
		 $\vee$ is cummutative & $+$ is commutative \\
		 $\wedge$ distributes over $\vee$ & $*$ distributes both ways over $+$ \\
		 $\vee$ distributes over $\wedge$ & ------ \\
		 0 is the minimum & P has an additive identity ``0" \\
		 1 is the maximum & P has a two--sided multiplicative identity ``1'' \\ \hline
    \end  {tabular}
\end  {center}

\noindent Although the usual $[0,1]\subset {\bf R}$ probabilities 
satisfy these conditions, they are only one possibility.  
At this stage, any ring will do, even a ring with non-commutative multiplication like the
quaternions. Actually, the fact that we have to explain interference effects
strongly suggests that we will need probabilities with an additive inverse.  
Plausibly also requiring scaling of probabilities by real numbers, we assume, at
this stage, that the probabilities of interest are real associative algebras
with unit.  Further restrictions are to come in section 3.

\section{Predicting frequencies}

    The exotic probabilities of the last section seem exotic mainly because 
we are immediately familiar with what, say, $P(b|a)=0.25$ means in terms of 
an experiment.  On the other hand, what is the predictive meaning of something 
like $(a\rightarrow b)=2+3i$?  To answer this, it is helpful to realize that
this problem already exists even in standard probability theory.  There is 
nothing in probability theory as such that tells us that probability $P(b|a)=0.25$ 
means $25{\%}$ should be expected in the corresponding frequency.  This must be 
deduced from additional assumptions.  In the standard probability case,
one considers $N$ copies of the situation where $a$ was known.  One
then observes that the probability that $b$ is true $n/N$ times peaks
at 0.25, and for any interval containing 0.25, the probability to be outside
the interval can be reduced as much as one wants by increasing $N$. Roughly 
speaking, the frequency meaning of standard
probabilities is fixed by the additional assumption that ``probability
zero propositions never happen.''  It may help to notice that, as Jaynes
points out\cite{ejaynes}, standard probability theory works equally
well on the interval $[1,\infty]$ rather
than $[0,1]$.  In this case, probability 4.0 would predict frequency
0.25 and one would be assuming that propositions with probability $\infty$ 
never happen.

    In the case of exotics, we cannot proceed quite as simply as in 
standard probability theory since, as will become clear, zero probability
propositions may sometimes be true anyway.  However, we can progress by 
assuming that $L$ contains a special subspace for which the standard
arguments will hold.  
Given $P$--probability $(L,\rightarrow)$, let $X$ be a measure space
and suppose that the free distributive
lattice on $X\times {\bf R}$ is a sublattice of $L$ \cite{nonmeasurablesets}.
We'll refer to the second component of $X\times {\bf R}$ as ``time'' and will 
often denote it as a subscript.  
For $A\subset X$, $A_t$ denotes $\bigvee_{a\in A} a_t$.
We will see below that frequency predictions follow if we assume
that $X$ has properties that one would expect of ``the state of the 
system.''  In particular, we assume that 
for any time $t$, $x_t\wedge y_t=0$ for
any $x,y\in X$ with $x\neq y$, meaning that ``the system can't be in two 
different states at the same time.''
Please note the clash of terminology 
with standard quantum theory where ``state space'' means a Hilbert space
and not just a measure space.

Given a state space $X$, and any fixed time $t$, we can relate probabilities to 
functions from $X$ to $P$.
For $a,b,c\in L$, let ``wave functions''
$\Psi_{a\rightarrow b}:X\rightarrow P$ be defined by

\begin{equation}
(a\rightarrow b\wedge \sigma_t) = \int_{\sigma} \Psi_{a\rightarrow b}
\end{equation}

\noindent for all measurable $\sigma\subset X$. 
Such functions are therefore related by 
\begin{equation}
\Psi_{a\rightarrow b\wedge c} = (a\rightarrow b)\ \Psi_{a\wedge b\rightarrow c}
\end{equation}
in general and 
\begin{equation}
\Psi_{a\rightarrow b\vee c} = \Psi_{a\rightarrow b} + \Psi_{a\rightarrow c}
\end{equation}
if $b\wedge c=0$.

In order to get real non-negative numbers from probabilities, we take $P$ to have
a square norm $\parallel\ \parallel:P\rightarrow {\bf R}^{0,+}$ satisfying
$\parallel p\ q\parallel=\parallel p\parallel\ \parallel q\parallel$ for $p,q\in P$.
Given this, we will show that, under certain conditions, 
\begin{equation}
{\rm Prob}_t(b|a)= {{\int_X \parallel \Psi^t_{a\rightarrow b}\parallel}\over{\int_X \parallel
\Psi^t_{a\rightarrow 1}\parallel}}
\end{equation}
is a probability in the ordinary sense.
When it doesn't cause confusion, we will
suppress the function name inside integrals as a notational convenience.
We may, for example, write 
\begin{equation}
{\rm Prob}_t(b|a)=
{{\int_{X} \parallel a\rightarrow b\wedge x_t \parallel}\over{\int_{X} \parallel a\rightarrow 1\wedge x_t \parallel}}.
\end{equation}
Note that probabilities like $(a\rightarrow b\wedge c\wedge x_t)$ are typically zero and, of course, 
$(a\rightarrow x_t)$ isn't equal to $\Psi^t_{a}(x)$.

    To derive properties of ${\rm Prob}_t$, note that 
\begin{equation}
{\rm Prob}_t(b\wedge c|a)=
{{\int_{X}\parallel a\rightarrow b\wedge c\wedge x_t\parallel}\over
{\int_{X}\parallel a\rightarrow x_t\parallel}}
\end{equation}
is equal to 
\begin{equation}
{ {\int_{X} \parallel a\rightarrow b\parallel\ \parallel a\wedge b\rightarrow c\wedge x_t
\parallel}
\over{\int_{X} \parallel a\rightarrow x_t\parallel} } *
{ {\int_{X} \parallel a\wedge b \rightarrow x_t\parallel} \over 
  {\int_{X} \parallel a\wedge b \rightarrow x_t\parallel} }
\end{equation}
and, rearranging and using 
$\parallel a\rightarrow b\parallel\ \parallel a\wedge b\rightarrow x_t\parallel =
\parallel a\rightarrow b \wedge x_t\parallel$, we have 
\begin{equation}
{\rm Prob}_t(b\wedge c|a)={\rm Prob}_t(b | a)\ {\rm Prob}_t(c|a\wedge b)
\end{equation}
as desired.  If we also knew that for $b\wedge c=0$, 
\begin{equation}
{\rm Prob}_t(b\vee c|a)={\rm Prob}_t(b|a) + {\rm Prob}_t(c|a)
\end{equation}
then we would have a complete standard probability theory and a
frequency meaning would follow as in the standard argument.  However, (12) is 
true if and only if
\begin{equation}
\int_X\parallel\Psi^t_{a\rightarrow b}+\Psi^t_{a\rightarrow c}\parallel =
\int_X\parallel\Psi^t_{a\rightarrow b}\parallel +
\int_X\parallel\Psi^t_{a\rightarrow c}\parallel
\end{equation}
which, in a Hilbert space setting, is equivalent to requiring $\Psi^t_{a\rightarrow b}$
and $\Psi^t_{a\rightarrow c}$ to be orthogonal.  
Thus, we've concluded that we can predict frequencies, but only for 
sublattices of $L$ for which $(12)$ holds.  This includes the sublattice 
$X$ at any fixed time and the sublattice of propositions associated with 
a Hermitian operator in the Hilbert space case.  

For example, suppose that we have an orthogonal set of functions 
$\{\phi_1,...,\phi_n\}$ in the Hilbert space $L^2(X)$ and suppose 
that $L$ contains the sublattice $B=\{b_1,b_2,\dots,b_n\}$ where
$b_i$ is the proposition ``$\phi_i$ is the best description of the 
system at time $t$.''  $B$ is a sublattice and (12) is satisfied because 
$<\phi_i,\phi_j>$ is zero for $i\neq j$ and so ${\rm Prob}_t$ on the sublattice
$B$ is therefore a probability theory in the ordinary sense and, for example
${\rm Prob}_t (b_j|\bigvee_{i=1}^{n}b_i)$ is the expected frequency that
$\phi_j$ is the best description of the system at time $t$ assuming that one of the 
$\phi_1,\phi_2,\dots\phi_n$ is optimal.

As another example, consider how we would describe a Stern--Gerlach 
experiment with quaternion probabilities and state space $X={\bf R}^3$.  At 
any time $t$ while the particle is heading towards the magnet, $X_t$ 
is a sublattice of $L$ and ${\rm Prob}_t$ is a standard probability theory
and predicts how often various subsets of $X$ are occupied.  
At a time $t'$ when the particle
has gone through the magnet and either gone up or down, $X_{t'}$ is 
also a sublattice and ${\rm Prob}_{t'}$ is also standard and predicts the
results of the experiment.  However, although $X_t\cup X_{t'}$ is a
a sublattice of $L$, we cannot conclude that either ${\rm Prob}_t$ or ${\rm Prob}_{t'}$
are standard probabilities because interference terms may prevent
(12) from being satisfied.  This is why exotic probabilities aren't 
eliminated by Bell's inequalities (see section 8).  
You can also see that this implies that the Stern--Gerlach experiment
is not a dynamical system.  If there was a function $f:X\rightarrow X$
such that a particle at $x_t$ always arrives at $f(x)_{t'}$, probabilities on
$X_{t}\cup X_{t'}$ would be determined by ${\rm Prob}_t$ and $f$.  In this
sense the Stern--Gerlach system is realistic but not deterministic.

Thus, we have found that exotic probabilities
can indeed acquire predictive power provided we introduce a ``state space''
within $L$ and a square norm on $P$.  Since the square norm
property $\parallel p\ q\parallel=\parallel p\parallel\ \parallel q\parallel$
is crucial, we conclude that probabilities must be
real associative algebras with a square norm. There are, however, only 
are only three such algebras: the reals, the complex numbers and the 
quaternions\cite{harvey}.  This means that particles may only be spin 0 or spin 1/2.
Since (12) is only prevented by ``interference terms'' we see that, in this sense,
``standard probability theory is restored in the classical limit.''

\section{More about state spaces}

    As pointed out in reference 4, modifying probability theory means that
we are free to simply assume that if a particle arrives at a point
$x_{t'}$ at a detecting screen in a two slit experiment, the particle was 
therefore somewhere in ${\bf R}^3$ at any previous $t\leq t'$.  In general, we assume that
\begin{equation}
x_{t'} = x_{t'} \wedge X_t
\end{equation}
for all $x\in X$, $t\leq t'$.
This has immediate implications.  For $t\leq t' \leq t''$, 
\begin{equation}
(X_t\rightarrow X_{t''})=(X_t\rightarrow X_{t'}\wedge X_{t''})=(X_t\rightarrow
X_{t'})(X_{t'}\rightarrow X_{t''})
\end{equation}
and if we also assume that probabilities are time invariant in the sense that
$(A_t\rightarrow B_{t'})=(A_{t+\tau}\rightarrow B_{t'+\tau})$ for any $A,B\subset X$,
$t,t',\tau\in {\bf R}$,
then $(X_t\rightarrow X_{t'}) = e^{\lambda (t'-t)}$ for some $\lambda\in P$.
This implies that 
$\Psi^{t'}_{X_t}(x) = e^{\lambda (t'-t)} \phi(x)$ for time independent 
$\phi:x\mapsto \int_\sigma (X_{t'}\rightarrow \sigma_{t'})$.
For those used to quantum mechanics, this may seem puzzling because, after assuming very 
little, we concluded that ``the system is in an energy eigenstate.''  What
if the system is, in fact in some other state?  If this question occurs to you, remember
that an exotic probability like $(X_t\rightarrow A_{t'})$ is only the best 
estimate that $A_{t'}$ is true given that $X_t$ is known.  If one knows some
additional facts $F$ about the system, one should instead calculate $(X_t\wedge F\rightarrow A_{t'})$.
Thus, our wave functions only represent what one knows about a system and can't 
be interpreted as ``the state of the system'' in any reasonable sense.  Different
observers will have different knowledge about a system and they may also describe 
a single system with different wave functions.  This means that if an 
observer does not know all the relevant facts about a system, their wave functions
may give incorrect predictions.  Of course, this is not a failure of exotic probability
theory 
any more than it is a failure of ordinary probability theory when the usual analysis of 
a die fails in the case of loaded die.  In both cases, the theories are successful to
the extent that relevant facts are known.  From the Bayesian view, the particular result 
above means that if one knows only that the system was somewhere in state space 
at time $t$, then the best description of the system at any later time is one of the 
energy eigenfunctions. 
   
    One last assumption completes what one intuitively means by a ``state space.''
Intuitively, if one knows the ``state'' $x_t$ at time $t\in {\bf R}$, then any previous
knowledge should be irrelevant.  In this sense, it is natural to assume 
\begin{equation}
(A_t\wedge x_{t'}\rightarrow B_{t''})=(x_{t'}\rightarrow B_{t''}).
\end{equation}
for any $t\leq t'\leq t''$, $A,B\subset X$, $x\in X$.  This assumption also has
immediate consequences.  For $A,B\subset X$, letting subscripts indicate time ordering
and using $\Psi^t_{a\rightarrow b}(x) = \Psi^t_a(x)\ (a\wedge x_t\rightarrow b)$,
\begin{equation}
(A_o\rightarrow B_n)=\int_{x\in X} \Psi^1_{A_o}(x)\ (A_o\wedge x_1\rightarrow B_n)=
\int_{x_1\in X} (A_o\rightarrow x_1)(x_1\rightarrow B_n)
\end{equation}
and, repeating the same argument,
\begin{equation}
(A_o\rightarrow B_n) = \int_{x_1,x_2,\dots,x_{n-1}} (A_o\rightarrow x_1)(x_1\rightarrow x_2)\dots(x_{n-1}\rightarrow B_n)
\end{equation}
for any sequence of intermediate times $t_1,t_2,\dots,t_{n-1}$.  We can refer to such 
an expression as a ``path integral.''  Note that this expression together with the 
definition of ${\rm Prob}$ means that ``paths interfere if they end at the same 
point in $X$.''  This is the exotic probability version of the ``which path'' principle of 
quantum mechanics.

\section{Definitions}

    Before continuing on to physics, let's collect the definitions so far and
establish some terminology.  For the rest of the paper, we assume lattices to be
distributive and to have minimum and maximum 
elements denoted ``$0$'' and ``$1$'' respectively. By 
a ``measure space,'' we always mean a measure space with a finite
real non-negative measure. 

Fix $P={\bf R}$, ${\bf C}$ or ${\bf H}$.  A $P$--probability is a lattice $L$ 
together with a function $\rightarrow:L\times L\rightarrow P$ satisfying
\begin{equation}
(a\rightarrow b\wedge c)=(a\rightarrow b)\ (a\wedge b\rightarrow c)
\end{equation}
for all $a,b,c\in L$ and satisfying
\begin{equation}
(a\rightarrow b\vee c)=(a\rightarrow b)+(a\rightarrow c).
\end{equation}
for all $a,b,c\in L$ with $b\wedge c=0$.

   Here are a few simple examples.
Let $L$ be the lattice $\{0,1\}$ and let $(a\rightarrow b)$ 
be $0$ if $b$ is the minimum and $1$ if $b$ is the maximum.  This 
is a $P$--probability.
   Given a lattice $L$, let $\phi:L\rightarrow P$ be some function satisfying 
$\phi(a\wedge b)=\phi(a)\phi(b)$
in general and $\phi(a\vee b)=\phi(a)+\phi(b)$ if $a\wedge b=0$.  Then $(a\rightarrow b)=\phi(b)$
makes $(L,\rightarrow)$ into a $P$--probability.  
   Let $L$ be a totally ordered lattice and let $(a\rightarrow b)$ be $1$ if 
$a\leq b$ and $0$ otherwise.  This is also a $P$--probability.  
   Given a $P$--probability $(L,\rightarrow)$ and a sublattice $M$ of $L$, 
let $l$ be an element of $L$.  We can then define a new $P$--probability $(M,\rightarrow_l)$
by letting $(a\rightarrow_l b)=(a\wedge l\rightarrow b)$ for $a,b\in M$.

   Following standard probability theory, we say that propositions $a,b\in L$
are independent if $(a\wedge q\rightarrow b)=(q\rightarrow b)$ for all $q\in L$
and this implies $(q\rightarrow a\wedge b)=(q\rightarrow a)(q\rightarrow b)$ 
as usual.
We say that subsets $A,B$ of $L$ are independent if $a$ and $b$ are independent
for all $a\in A$ and $b\in B$.

    Given a $P$--probability $(L,\rightarrow)$, we can 
define the product of independent sublattices $M$ and $N$ of $L$.  
Letting $(M\times N,\rightarrow_\times)$ be defined by 
\begin{equation}
(m,n)\rightarrow_\times(m',n') = (m\rightarrow m')(n\rightarrow n').
\end{equation}
This defines a $P$--probability, even if $P$ is not commutative.

    Let $X$ be a measure space and let ${\cal F}X$ be the free lattice on $X\times {\bf R}$
subject to
\begin{equation}
x_t\wedge y_t = 0
\end{equation}
for all $x,y\in X$, $x\neq y$, $t\in {\bf R}$ and 
\begin{equation}
x_{t'}=x_{t'}\wedge X_t
\end{equation}
for $x\in X$ and times $t\leq t'$.
A $P$--probability $(L,\rightarrow)$ is said to ``have a state space $X$'' if 
${\cal F}X$ is a sublattice of $L$ and if 
\begin{equation}
(A_t\wedge x_{t'}\rightarrow B_{t''})=(x_{t'}\rightarrow B_{t''})
\end{equation}
for all times $t\leq t'\leq t''$ for all subsets $A,B\subset X$ and for all $x\in X$.

\section{A simple interferometer}

    To exercise our ideas so far, let's 
analyze the interferometer shown in figure 1 in some detail. Although one is 
instinctively shy at first, we are free to use simple language
to describe what happens as if the particle was a marble.

\begin{figure}[htbp]
\begin{center}
  \leavevmode
  \epsfig{file=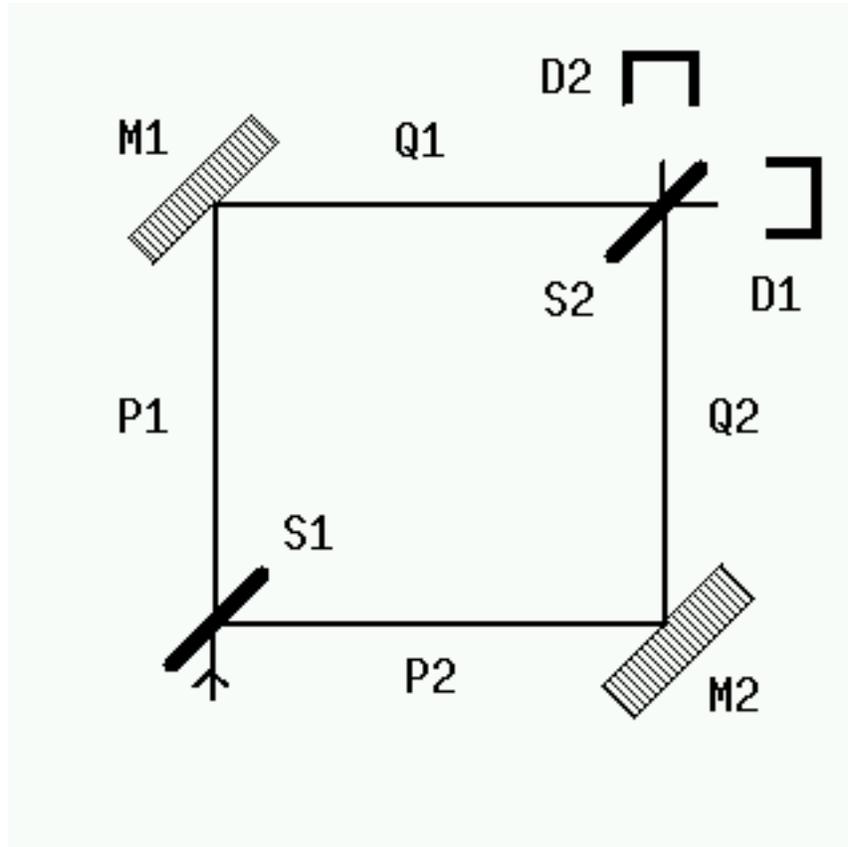}
  \caption{ A simple interferometer where a particle enters as indicated and encounters
  a beam splitter ($S_1$), a mirror ($M_1$ or $M_2$) and a second beam splitter ($S_2$) ending up either
  in detector ($D_1$) or ($D_2$).}
  \label{fig:testfig1}
\end{center}
\end{figure}

\noindent Working within a ${\bf C}$--probability with state space $X={\bf R}^3$,
we can say that a particle 
hits $S_1$ and either goes on the $P_1$ branch
or the $P_2$ branch.  After hitting either mirror $M_1$ or $M_2$, the
particle is on the $Q_1$ or the $Q_2$ branch respectively.  The particle
will hit $S_2$ and will end up in either detector $D_1$ or in detector 
$D_2$.  Experimentally, one surprisingly finds that particles always 
end up in $D_2$.  Letting ``$e$'' informally denote the experimental 
arrangement, we would like to calculate $(e\rightarrow D_1)$ and $(e\rightarrow D_2)$.
Since $D_j$ implies both $P_1\vee P_2$ and $Q_1\vee Q_2$, we have
$(e\rightarrow D_j)=(e\rightarrow(P_1\vee P_2)\wedge(Q_1\vee Q_2)\wedge D_j)$.
Using $P_1\wedge P_2 = Q_1\wedge Q_2 = 0$, we mechanically apply axioms to 
produce
\begin{equation}
(e\rightarrow D_j) = \sum_{n,m=1}^2 (e\rightarrow P_n)(e\wedge P_n\rightarrow Q_m)(e\wedge P_n\wedge Q_m\rightarrow D_j).
\end{equation}
Since $P_1$ is equivalent to a point in $X$, previous knowledge is irrelevant and 
we have $(e\wedge P_n\rightarrow Q_m)=(P_n\rightarrow Q_m)$.  
We also clearly want to assume that the particle can't hop the rails, in other
words we assume that $(P_n\rightarrow Q_m)$ is zero unless $n=m$.
This causes one of the sums to disappear giving
\begin{equation}
(e\rightarrow D_j) = \sum_{n=1}^2 (e\rightarrow P_n)(P_n\rightarrow Q_n)(Q_n\rightarrow D_j)
\end{equation}
This result is not surprising, but the point to focus on is that the result
follows rigorously from the exotic probability axioms with natural assumptions
given the marble--like picture of what is happening.

To proceed further, we have to define what happens at the mirrors and the beam 
splitters.  Naturally, in either this case or in standard quantum theory, what
one means by ``a mirror'' and ``a beam splitter'' has to be put in by hand.  
In the ideal case, what one means by a ``mirror'' is that complex probabilities
of particle bouncing off of it pick up a factor of $i$.  A good experimentalist would
naturally test this assumption in other measurements.  Similarly, the beam splitters
multiply probabilities by a factor of $i$ when there is a ``bounce.'' Thus, 
$(e\rightarrow P_2)=i*(e\rightarrow P_1)$,
$(Q_1\rightarrow D_2)=i*(Q_1\rightarrow D_1)$, $(Q_2\rightarrow D_1)=i*(Q_2\rightarrow D_2)$, 
and $(P_1\rightarrow Q_1)=(P_2\rightarrow Q_2)$ and so
$(e\rightarrow D_1)=0$ as expected.  

Suppose now that the interferometer is such that a device could be attached to 
$M_1$ such that it registered ``hit'' or ``nohit'' depending on whether the
particle struck $M_1$ or not.  Experimentally the results are different and about
half the particles go into $D_1$.
In quantum theory, one says that this is due to the ``which path'' principle.  The
two paths ending in $D_1$ no longer interfere because ``you can tell which path 
was taken.''
You can see that this result also follows mechanically with exotic probabilities.  
In the described situation,
${\bf R}^3$ is evidently not a sufficient state space and one
should use at least ${\bf R}^3\times\{{\rm hit},{\rm nohit}\}$.  In this case, one can
explicitly calculate that the 
interference is lost because two paths ending in $D_1$ no longer end at the same
point in the state space. One can also calculate that if the device detecting whether $M_1$ is hit
works so poorly that $\{{\rm hit},{\rm nohit}\}$ are independent of $Q_1$ and $Q_2$,
then the interference effect is entirely restored\cite{mpl2}.

Note the difference with standard quantum theory.
Quantum mechanics has no problem with this interferometer in the sense that  
the wave equation can be solved for any desired input wave packet.  Of course, 
no one wants to do this, especially to get such simple results.  This 
explains the popularity of the ``which path'' principle even though it is 
not completely clear what it means or how it follows from the fundamental 
wave equation.  This is analogous to doing probability theory knowing
the diffusion equation but not knowing Kolmogorov's axioms.
In exotic probabilities, on the other hand, both a rigorous version of 
the ``which path'' principle and any wave equation are consequences of the
underlying exotic probability theory.

\section{Exponential Decay}

   The interferometer from the previous section suggests that exotics may be 
particularly helpful in situations where one wants predictions
which are independent of details of initial wave functions and potentials. 
``Exponential decay'' provides simple examples of such situations and also 
brings up one of the lesser known mysteries of quantum theory.  
Consider a system such as a ${\rm Co}^{60}$ nucleus or a muon which may 
decay irreversibly.  Given such a system, if the probability for a decay 
within a time interval $t$ only depends on $t$ and not on the history of 
the system, then a familiar argument in probability theory implies that 
the probability density for decay is exponential.  Quantum mechanics, however, does 
not generally predict this\cite{sakurai} and so it would
seem that for such
non--exponential systems, the assumption that they decay independent of their
history is not correct.
As with other paradoxes\cite{santafe}, we can resolve this by realizing that
the physical assumptions are correct; the problem is caused by probability 
theory itself.
Applying the physical assumptions to exotic probability theory instead, we suppose that in a $P$--probability with state space $X$,
$(A_t\rightarrow B_{t'})=(A_{t+\tau}\rightarrow B_{t'+\tau})$ for all $t,t',\tau\in {\bf R}$.
Suppose also that $X$ contains a subset $\alpha$ whose complement $\beta$ is a ``trap''
in the sense that $\beta_t$ implies $\beta_{t'}$ for any $t\leq t'$.  This 
means that $\alpha_{t'}$ implies $\alpha_{t}$ for any $t\leq t'$ also.  With arguments similar to 
those in section 4, we find 
$(\alpha_0\rightarrow \alpha_t) = e^{\lambda t}$,
$(\beta_0\rightarrow \beta_t) = 1$,
$(\alpha_0\rightarrow \beta_t) = a\ (1-e^{\lambda t})$,
and $(\beta_0\rightarrow \alpha_t) = 0$
for some $\lambda\in P$ and $a\in {\bf R}$.  Although the exotic probabilities are simple 
exponentials, this isn't preserved in the predicted frequencies.  The ordinary 
probability to remain free for time $t$ is
\begin{equation}
{\rm Prob}(\alpha_t|\alpha_0) = {{\int_\alpha\parallel\alpha_0\rightarrow x_t\parallel}\over
{\int_\alpha\parallel\alpha_0\rightarrow x_t\parallel + \int_\beta\parallel\alpha_0\rightarrow x_t\parallel}}
\end{equation}
and, using 
$\int_\alpha\parallel\alpha_0\rightarrow
x_t\parallel=\parallel\alpha_0\rightarrow\alpha_t\parallel\int_\alpha\parallel\alpha_t\rightarrow x_t\parallel$ and
$\int_\beta\parallel\alpha_0\rightarrow x_t\parallel = 
\parallel\alpha_0\rightarrow\beta_t\parallel\int_\beta\parallel\alpha_0\wedge \beta_t \rightarrow x_t\parallel$,
we have
\begin{equation}
{\rm Prob}(\alpha_t|\alpha_0)={ {1}\over{1+k(t)\parallel e^{-\lambda t} - 1\parallel} }
\end{equation}
where
\begin{equation}
k(t) = a^2\ {{\int_\beta\parallel\alpha_0\wedge\beta_t\rightarrow x_t\parallel}\over{\int_\alpha\parallel\alpha_t\rightarrow
x_t\parallel}}.
\end{equation}
For small $t$ and assuming that $\lambda$ is real an negative, 
${\rm Prob}(\alpha_t|\alpha_0)$ will decrease more slowly than $1-2\lambda t$.  
If we also know that $\alpha_0$ and $x_t\in \beta_t$ can be taken to be independent for 
sufficiently large $t$, then we say that the system is ``forgetful.'' In this case, $k(t)$ 
is asymptotically constant
and ${\rm Prob}(\alpha_t|\alpha_0)$ will be exponential for large times.
Such deviations from exponential decay have only recently been observed
experimentally\cite{Wilkinson}.

    The examples of the last two sections show the usefulness of applying exotic probability 
theory directly as opposed to solving a PDE.  This sort of reasoning is mostly missing in 
standard quantum theory.

\section{Bell's inequalities}

    Bell's well known analysis of the spin version of the Einstein--Podolsky--Rosen
experiment\cite{bell} is  
almost universally summarized as showing that local realistic theories
are incompatible with the predictions of quantum mechanics and are 
therefore wrong.  One might then expect that exotic probabilities
would be ruled out by Bell because they are ``realistic'' in the state space
sense. 
Bell's analysis, however, does not follow once we modify probability theory.
To see the problem, you only have to notice that the first step in Bell's
analysis
assumes that $P(M_{t'}|e)=P(M_{t'}\wedge \Lambda_t |e)$ and
\begin{equation}
P(M_{t'}\wedge \Lambda_t|e) = \int_{\lambda\in\Lambda} P(M_{t'}\wedge\lambda_{t}|e) = 
\int_{\lambda\in\Lambda} P(\lambda_{t}|e)P(M_{t'}|e\wedge \lambda_{t})
\end{equation}
for initial setup $e$, final measurement $M_{t'}$ and assuming that
the final results are determined by some ``hidden variable'' 
$\lambda\in\Lambda$ at some time $t$ during the flight from decay 
to detectors.  As pointed out
in section 3, equation 33 fails to hold in general due to ``interference terms''\cite{pl}.  
In fact, Bell has shown exactly that if one wants local realism one must modify
probability theory.  Ironically, the standard summary of his results 
gives the opposite impression.

   Over the years, there have been more than twenty variations on Bell's 
result each with a different experimental arrangement and each concluding
that local realistic theories are impossible.  Bell's result and two of 
the more well known variations are considered in reference 3 in some detail
and are shown not to eliminate exotic probabilities.  There has also been
an increasing tendency to refer to Bell and similar results as ``non--local''
effects because they cannot be explained by local correlations\cite{pl}.  The point
is, however, that if one has the wrong probability theory, one may also have the
wrong notion of what is just a correlation.  Within exotic probability theory,
we expect that Bell's results are just correlations in the new probability theory.
It's helpful to think of a classical experiment where one cuts a penny into
a heads half and a tails half and mails one half penny to house $A$ and the
other half to house $B$.  The 
results at the two houses are correlated, but nothing travels between them 
to insure the proper results.  One therefore expects that there is nothing
that one can do at house $A$ to affect the fact that, at house $B$, one will
find heads 50\% of the time and tails 50\% of the time.  The same
holds true in the EPR experiment.  
The results at one end of the experiment are
50\% spin up and 50\% spin down independent of the magnet orientation
nothing that happens on the other side can affect this.
  
\section{Time evolution}

    Given some initial knowledge such as $A_t$ with $A\subset X$, the exotic probability
to arrive at some $B\subset X$ at some later time $t''$ is given by 
\begin{equation}
(A_t\rightarrow B_{t''}) = \int_{x\in X} (A_t\rightarrow x_{t'})(x_{t'}\rightarrow B_{t''})
\end{equation}
for any time $t'$ with $t\leq t'\leq t''$.
This is called the Chapman--Kolmogorov equation in the probability literature.
In the complex case with state space ${\bf R}^d$, one can either follow reference 4
or Risken\cite{risken} to conclude 
that for small $\tau\in {\bf R}$ and small $z\in X$, $(x_t\rightarrow (x+z)_{t+\tau})$ is given by 
\begin{equation}
{ {1}\over{(2\pi\tau)^{d/2}\sqrt{{\rm det}(\nu)} } }
{\rm exp}(-\tau[{{1}\over{2}}({{z_j}\over{\tau}} - \nu_j)\nu_{jk}^{-1}
({{z_k}\over{\tau}} - \nu_k)+\nu_o])
\end{equation}
where $\nu_o$, $\nu_j$ and $\nu_{jk}$ are moments of the time derivative of 
$\omega(x,z,\tau)\equiv(x_t\rightarrow (x+z)_{t+\tau})$
defined by complex functions 
$\nu_o(x)\equiv\int_X \omega_\tau(x,z,0)$, $\nu_j(x)\equiv\int_X \omega_\tau(x,z,0)z_j$,
$\nu_{jk}(x)\equiv\int_X\omega_\tau(x,z,0)z_j z_k$.  This is a central--limit--theorem--like 
phenomena where the details of the unknown function $(x_t\rightarrow (x+z)_{t+\tau})$ are smoothed
over and only a dependence on it's lowest moments survives.  Identifying $z_j/\tau$ as the velocity, 
equation 35 is equivalent, for example, to the Schrodinger equation in ${\bf R}^3$ identifying
$\nu_o=-ieA_o$, $\nu_j={{e}\over{m}} A_j$ and $\nu_{jk}=(i/m)\delta_{jk}$.
Similarly, quaternion probabilities in result in
the Dirac equation\cite{srinivasan1,srinivasan2}. 
These arguments need to be made into proofs, but
there is also a mystery as to why only parts of the available moments seem to be 
used by nature.  Why, for instance, must $\nu_j$ be purely real in ${\bf R}^3$?

\section{Comparison with quantum theory}

   In standard quantum theory, the state of the system is a ray in 
a Hilbert space.  To define such a theory one must define a Hilbert space
and a complete set of mutually commuting self-adjoint operators to serve 
as observables.  In addition, one chooses a Hamiltonian and labels the 
states in the Hilbert space by irreducible representations of the 
Hamiltonian's symmetry 
group.  For Hamiltonians invariant under the Lorentz group, states
have spin and four--momenta. Time evolution is a one parameter semigroup given by
the Hamiltonian operator.  If ``mixed states" occur, they must be described
by density matrices. Quite a bit of functional analysis must be 
understood to define this precisely.

   In an exotic probability theory, on the other hand, 
the state of the system is a point in a measure space $X$.  To define the 
theory, one simply chooses $X$ and picks ${\bf R}$, ${\bf C}$ or ${\bf H}$.
Particles are not thought of as having momentum or spin, or any other 
internal structure.  The only thing that a particle can do is to be 
somewhere.  This is all that is required, however, because experiments which
measure things like momentum and spin are always ultimately measuring position.
Wave functions have the same status as densities do in Bayesian theory.
People with different knowledge about a system will, in general, use 
different wave functions.  Those who have more knowledge can expect better
predictions. Situations requiring ``mixed states'' in quantum theory are described
by the same exotic theory without modification\cite{mpl2} and, similarly, there is
no sensible concept of ``being in a mixed state.'' 
Rather than choosing a Hamiltonian, one notes that wave functions are
propagated in time by the unknown $(x_t\rightarrow x'_{t'})$.  In typical state 
spaces this propagation
obeys a PDE which depends only upon the lowest moments 
of $(x_t\rightarrow x'_{t'})$ and these moments are identified with the vector potential 
and metric tensor.  The relevant moments can either be measured experimentally with 
test particles or computed with some external theory like Maxwell's equations.
One does not assume Lorentz or gauge invariance to get these results.

\section{Implications for the rest of physics and open questions}

    Physical theories are thought to be quantum theories in only
in a somewhat general sense.  The successful predictions of quantum
mechanics, must, of course, be reproduced, but this is not taken to mean
that any theory must literally satisfy the axioms of quantum theory.  
There is, however, an independent reason why physical theories must
be precisely exotic probability theories.  The results of section two
and three indicate that any theory which assigns likelihoods
to pairs of propositions from a distributive lattice must exactly 
be an exotic probability theory or must violate one of our two Cox
conditions or must fail to reduce to standard probability theory 
when predicting frequencies.  Physical theories are constrained by 
the results here just as alternatives to standard probability theory
are constrained by Cox's original arguments.  The implications of this
raise many questions about how this should be done for the rest of 
physics.

   In the case of field theory, Srinivasan has pioneered application
of exotic probabilities to quantum field theory by calculating the 
Lamb shift in a quaternionic version of canonical quantization.  His
results agree with QED without any renormalization
procedure.  
In addition to Srinivasan's approach it is clear in a very 
simple sense that electrons 
must emit photons because the vector potential remains unknown even when the 
electromagnetic field has been measured. Even in the case of a single 
electron, one must therefore sum over the various possible gauge equivalent
vector potentials.  One has no choice but to predict that an electron
will have various possible motions and these will be correlated with 
various possible vector potentials.  It is reasonable to expect that 
this simple effect should fit naturally in the framework of a complete
field theory.  This, however, has not been done.  Also, similar 
considerations hold for the metric tensor and weighted sums over
various possible metric tensors must similarly be finite.  Does this then mean
that one could calculate gravitational radiation?  

    Exotic probability theories are much more restrictive than
quantum mechanics in the sense that the form of the vector potential
and metric tensor is already determined by the choice of state space and
probability.  Since the choice of probability seems to be fixed by
spin, one apparently only has the state space left to explain things
like other gauge theories besides QED. Can Yang-Mills theories be
formulated as exotic probability theories, and, if so, with what
state space?

    Other questions arise if we sketch the general procedure for 
finding a PDE for wave functions.  The basic theory here is formulated
with a state space $X$ only assumed to be a measure space.  Assuming
that $X$ also has a topology, consider a point $x$ in an open set
$O\subset X$.  One assumes that a time difference $t'-t$ can be 
chosen such that $(x_t\rightarrow x'_{t'})$ is negligible for
$x'$ outside of $O$.  In addition, we suppose that $O$ can be 
chosen such that $(x_t\rightarrow x'_{t'})$ can be 
approximated by a function of only $x'-x$ and $t'-t$.  Given this, 
the path integral within $O$ collapses to a convolution and this
can be inverted with a Fourier transform resulting in a kernel 
depending only on the lowest moments of the time derivative of
$(x_t\rightarrow x'_{t'})$ as in section 9.  Another way to 
think about this is to consider the ring of $P$--valued 
functions on $O$ with pointwise addition and convolution 
as multiplication.  In this case, we assume that these rings have units 
$K_t$ in a ``Dirac 
sequence'' sense\cite{lang} ${\rm lim}_{t\rightarrow 0}\ K_t * f = f$ and
${\rm lim}_{s,t\rightarrow 0}\ K_s * K_t = K_{s+t}$.  This can be 
solved by considering a slight generalization of the standard quadratic
form: a function $q:V\rightarrow P$ where $q(x+y)-q(x)-q(y)=b(x,y)$ for
a symmetric bilinear $b:V\times V\rightarrow P$. Then 
$K_t:x\mapsto e^{q(x-a)/t}/I_t$, $a\in V$, $I_t=\int_{O} e^{q(x)/t}$
provides solutions.  As mentioned in section 9, this raises the question 
of why only certain of these $K_t$ are seen in nature.  There is also 
the question of what exactly must be assumed about $X$ since, 
besides a topology, we 
only seem to need subtraction of nearby points in $O$.  For instance, 
it is perhaps interesting to remove geometry entirely by allowing
any multiplication on ${\rm Hom}(O,P)$
which forms a ring with pointwise addition and has a unit in the Dirac sequence 
sense.

    Although Srinivasan has worked in field theory directly, simple multi--particle
systems have not been done with exotic probabilities. In particular, what is
the relationship between spin and statistics for exotic probabilities?
This seems likely to be interestingly different than in standard field theory.

   Although the time parameter in exotics seems essential once the state
space axioms are introduced, this does not mean that exotics are
nonrelativistic.  ``Time'' in the complex ${\bf R}^4$ theory, for 
example, can be interpreted as the proper time or path length
parameter.  One suspects however,
that ``time'' is really the order in which one discovers
facts about the system rather than anything more intrinsic.  
In this case, one might expect that automorphisms of the
time parameter should
result in equivalent theories with modified moments of 
$(x_t\rightarrow x'_{t'})$.  Is this correct and, if so, what
are the consequences of invariance under time automorphisms?

The fact that the vector potential appears as the first moment 
of the time derivative of $(x_t\rightarrow {x'}_{t'})$ suggests that Maxwell's 
equations should describe complex or quaternionic vector 
potentials.  Are there complex and quaternionic versions of Maxwell's
equations and, if so, are it's classical predictions correct?

   The whole area of ``Bayesian Inference'' in ordinary probability theory
is based on the idea that one can used Bayes theorem (which also follows in 
exotics) to systematically improve probabilities based on ``prior'' knowledge.
It is clear that the same thing should be possible with exotic probabilities.
In the standard Bayesian case, this is often based on the maximum entropy 
principle. The issue, then, is how to do Bayesian inference and is there an
analogue of maximum entropy?

\section{Summary}

    Exotic probability theories as described here appear to be the  
only generalization of probability theory consistent with the basic
Bayesian framework.  In addition to standard probability theory, we
find that three exotic copies are possible where probabilities are real, complex or
quaternion valued respectively.  Although the exotic theories are substantially simpler
that quantum mechanics both conceptually and mathematically, they nevertheless
give the same predictions as standard quantum theory.  These theories 
constrain physical theories in the same sense that
Cox's original arguments constrain possible alternatives to standard
probability theory.  The implications of this beyond basic quantum
theory are mostly unexplored, but we have attempted to at least 
formulate some fundamental open questions where new insights are needed.

\section{Acknowledgement}

I am grateful to Robert Kotiuga and Tom Toffoli for helpful discussions.


\begin{thebibliography}{10}
\bibitem{nonmeasurablesets} Assuming that $X\times T$ is a sublattice of $L$ implies 
that probabilities such as $(a\rightarrow N_t)$ can occur where $N$ is a non--measurable
set.  This does not cause problems because we only assume that 
$(a\rightarrow b\vee c)=(a\rightarrow b)+(a\rightarrow c)$ for pairs $(b,c)$ and not
for countable subsets $b_1,b_2,\dots$ with $b_i\wedge b_j=0$ for all $i\neq j$.  ``Measure
space" in this paper refers to a measure space with a finite measure.
\end{thebibliography}

\end{document}